\documentclass[twocolumn,twocolappendix,times,
]{aastex631}

\begin{document}

\title{Cosmology and Astrophysics with the Diffuse eRASS1 X-ray Angular Power Spectrum}

\correspondingauthor{Erwin T.\ Lau}
\email{erwin.lau@cfa.harvard.edu}

\author[0000-0001-8914-8885]{Erwin T.\ Lau}
\affiliation{Center for Astrophysics, $\vert$ Harvard \& Smithsonian, 60 Garden Street, Cambridge, MA, 02138, USA}
\affiliation{University of Miami, Department of Physics, Coral Gables, FL 33124, USA}

\author[0000-0003-0573-7733]{\'Akos Bogd\'an}
\affiliation{Center for Astrophysics, $\vert$ Harvard \& Smithsonian, 60 Garden Street, Cambridge, MA, 02138, USA}

\author[0000-0002-6766-5942]{Daisuke Nagai}
\affiliation{Department of Physics, Yale University, New Haven, CT 06520, USA}

\author[0000-0002-1697-186X]{Nico Cappelluti}
\affiliation{University of Miami, Department of Physics, Coral Gables, FL 33124, USA}

\author[0000-0002-1706-5797]{Masato Shirasaki}
\affiliation{ National Astronomical Observatory of Japan (NAOJ), National Institutes of Natural Science, Mitaka, Tokyo 181-8588, Japan }
\affiliation{The Institute of Statistical Mathematics, Tachikawa, Tokyo 190-8562, Japan }

\begin{abstract}
The recent tension in the value of the cosmological parameter $S_8 \equiv \sigma_8(\Omega_M/0.3)^{1/2}$, which represents the amplitude of the matter density fluctuations of the universe, has not been resolved. In this work, we present constraints on $S_8$ with the X-ray angular power spectra of clusters and groups measured with the half-sky map from the eROSITA All Sky Survey data release 1 (eRASS1). Thanks to the extensive sky coverage of eRASS1, it achieves unprecedented precision compared to previous power spectrum measurements. Using a well-calibrated, physical halo gas model that includes astrophysics of feedback and non-thermal pressure support, we obtain $S_8 = 0.80^{+0.02}_{-0.01}$ with 1$\sigma$ uncertainty that is competitive against other cosmological probes. Our derived $S_8$ value is smaller than the primary CMB measurements from {\em Planck}, but still consistent to within $1\sigma$.  We also obtain constraints on the astrophysics of feedback, non-thermal pressure, equation of state in cluster cores, and outer boundaries of gas in clusters and groups. We discuss how additional X-ray observations, and cosmological surveys in microwave and optical, will further improve the cosmological constraints with the angular power spectrum. Our work demonstrates that the angular power spectrum of clusters and groups is a promising probe of both cosmology and astrophysics.
\end{abstract}


\section{Introduction} \label{sec:intro}

As the largest and most massive virialized structures in the Universe, galaxy clusters uniquely probe both astrophysics and cosmology. Cosmological constraints from galaxy clusters are usually obtained via cluster abundance or cluster gas fraction \citep[e.g., see][for review]{allen_etal11}.  Galaxy cluster surveys in the past decade have yielded cosmological constraints with increasing precision \citep{vikhlinin_etal09, mantz_etal10, benson_etal13, planck_sz_counts, des_y1_clusters21, spt_2019, spt24}. 
 
However, the cosmological constraints derived from galaxy clusters, as well as other low redshift cosmological probes, such as cosmic shear, galaxy clustering \citep{kids450, kids1000_lens, des_y3}, showed tension with the cosmic microwave background (CMB) measurements with {\em Planck} \citep{planck2018} on the $S_8$ parameter \citep{planck2018} at the 1-3$\sigma$ level. The parameter $S_8 \equiv \sigma_8(\Omega_M/0.3)^{1/2}$ combines two key cosmological parameters: $\sigma_8$ and $\Omega_M$ that represent the fluctuation and the amount of matter in the Universe, respectively. The origin of this ``$S_8$" tension is still uncertain. It could be due to non-linear physics that suppresses the matter power spectrum at late cosmic times \citep{amon_efstathiou22}, such as baryonic physics that bias lensing-based cosmological measurements at low redshifts \citep{mccarthy_etal23}. Or, the tension could be due to new physics, such as decaying dark matter \citep{tanimura_etal23}. Adding more to the mystery, recent cluster abundance measurement from the eROSITA All Sky Survey \citep{eRASS1} yields $S_8$ constraint that is consistent with {\em Planck} \citep{eRASS1_cluster_abund} but higher than the value obtained from other low-redshift probes at the $3\sigma$ level. 

The angular power spectrum of galaxy clusters and groups offers a new and alternative way of constraining cosmological parameters, such as $S_8$. It measures the correlations of signals coming from cluster- and group-size halos, over ranges of halo masses and redshifts at given angular separations on the sky. The thermal Sunyaev-Zeldovich (tSZ) angular power spectrum of clusters and groups is predicted to be a very sensitive probe to $\sigma_8$. Specifically, the amplitude of the tSZ cluster power spectrum scales as $C_\ell \sim \sigma_8^8$ \citep{komatsu_seljak02, bolliet_etal18}, making it an excellent probe of $S_8$. As the angular power spectrum can be measured directly at the map level, it does not require the galaxy cluster mass calibration. Thus, angular power spectrum measurements offer an independent way of obtaining cosmological constraints that are not subjected to the systematics associated with cluster masses in cluster abundance experiments. 

The angular power spectrum of galaxy clusters has been measured previously in the microwave band via the tSZ effect with {\em Planck} \citep{planck_sz_power}, which was limited by {\em Planck}'s low angular resolution ($\sim 10$ arcminutes), as well as systematic effects due to dust and star-forming galaxies. On the other hand, the angular power spectrum measurements in X-ray have been measured on surveys on smaller sky patches, such as those from {\em Chandra} \citep{cappelluti_etal12, kolodzig_etal17, kolodzig_etal18}, and {\em Suzaku} \citep{yu_etal2022}. These measurements have small-sky coverage, leading to large uncertainties in the measured power due to cosmic variance. Until recently, ROSAT was the only X-ray observation with a large sky coverage \citep{snowden_etal97}, but with its limited sensitivity, it had only led to an upper limit on $\sigma_8$ \citep{diego_etal03}. Some of the systematics in the {\em Planck} tSZ power spectrum measurements can be taken out by cross-correlating with ROSAT, but the uncertainties in the derived cosmological constraints remain relatively large compared to other cosmological probes \citep{hurier_etal15} due to limitations of both instruments.  

The recent observations from the eROSITA all-sky survey are promising as it has large sky coverage with improved sensitivity that is comparable to XMM-{\em Newton}. With the eROSITA Final Equatorial-Depth Survey \citep[eFEDS;][]{eFEDS}, we measured the X-ray angular power spectrum of extended sources and show that the measurements are consistent with the prediction of the halo model \citep{lau_etal23}. However, the $140 \ \rm{deg^2}$ sky area of eFEDS, which only covers $3\%$ of the total sky, is not large enough to obtain competitive cosmological parameters, since the cosmic variance of the power spectrum scales inversely with the square root of the sky coverage. 
The recently released half-sky X-ray data from the eROSITA-DE All Sky Survey \citep[eRASS1,][]{eRASS1}, with much larger sky coverage, provide us with an opportunity to constrain cosmology and cluster astrophysics with the X-ray angular power spectrum with higher precision.  Our angular power spectrum measurement, which is dominated by X-ray clustering signal inside the same halo (i.e., the one-halo term), is complementary to the two-point correlation function measurements of resolved eRASS1 clusters in \citet{eRASS1_cl_clustering}, whose clustering signal is mostly coming from two-halo clustering. 

In this paper, we present measurements of the X-ray angular power spectrum from the eRASS1 X-ray surface brightness map. We derive constraints on the feedback and non-thermal pressure of the gas in cluster- and group-size halos, as well as constraints on $S_8$. The paper is organized as follows.  In Section~\ref{sec:data}, we present and describe the analysis of the eRASS1 data.  In Section~\ref{sec:model}, we describe our model of the X-ray angular power spectrum model. In Section~\ref{sec:results}, we present our comparison between our model and observations. We provide a discussion of the results in Section~\ref{sec:discussion} and conclusions in Section~\ref{sec:summary}. 

\section{Data Reduction and Analysis}
\label{sec:data}

\subsection{Construction of the eRASS1 half-sky map}\label{sec:erass_map}

\begin{figure*}
    \centering
    \includegraphics[width=0.99\textwidth]{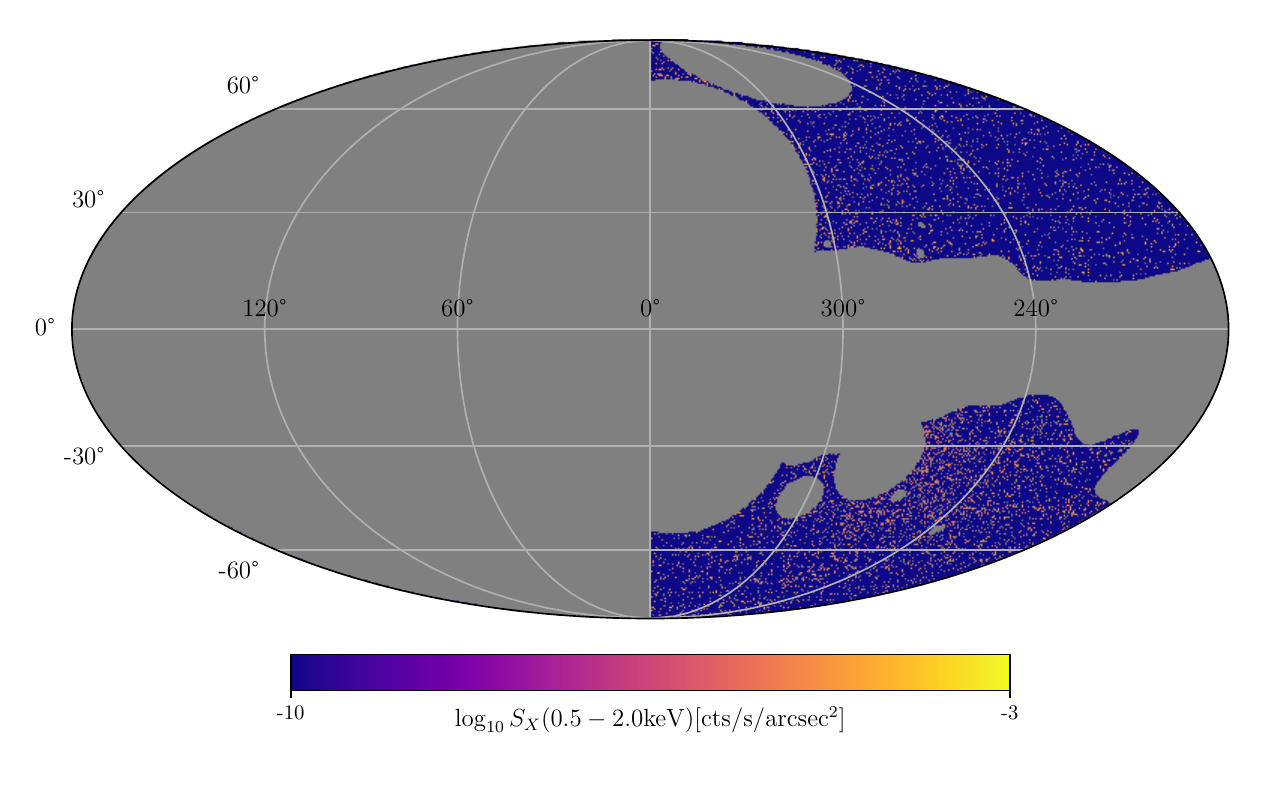}
    \caption{The eRASS1 X-ray photon map in count rate per square arcsecond in the $0.5-2.0$~keV band used in our analysis. The Galactic plane, point sources from the eRASS1 source catalog, eROSITA bubbles, and the LMC and SMC masked out. Note that we do not have access to the data in the eastern half of the map.  }
    \label{fig:eRASS1_map}
\end{figure*}

We construct the half-sky map using all the 2447 tiles with uniform area of $3.6\deg \times 3.6\deg$, from the eRASS1 scanned sky maps that are publicly available from {eROSITA}-DE DR1.\footnote{\url{https://erosita.mpe.mpg.de/dr1/}} For each tile, we construct the image from the event fits file using the {\tt evtool} command from the eROSITA Science Analysis Software System pipeline\footnote{\url{https://erosita.mpe.mpg.de/dr1/eSASS4DR1/}} (eSASS) version 010. When constructing the image map, we select photons with energy between 0.5 and 2.0~keV, discarding the data from the two detectors (TM5 and TM7) that had been contaminated with light leakage \citep{eROSITA}, with {\tt flag=0xE000F000} and {\tt pattern=15}. The corresponding exposure maps are generated using the {\tt expmap} command, with flags {\tt withdetmaps=YES}. We also set {\tt withweights=YES} and {\tt withvignetting=YES} to correct for the number of detectors used and vignetting of the telescope, respectively. Due to the scanning strategy of eROSITA, the exposure is not spatially uniform. It ranges from around $100$ s and the ecliptic equator to $10$ ks at the ecliptic poles. 

The image maps are then stitched together onto a half-sky map with HEALpix pixelization \citep{healpix}, with $N_{\rm side}=8192$, corresponding to a pixel size of $\sim 25.7\arcsec$. The half-sky map is generated by first stitching the tile maps in HiPS (Hierarchical Progressive Surveys) format, with a modified version of the eROSITA-specific HiPS map generation code, \footnote{\url{https://github.com/jeremysanders/ero_hips_gen}} and then converting the HiPS map into HEALpix format. \footnote{\url{https://gist.github.com/tboch/f68cd1bb1529d8ac12184b40e54ba692}}

\subsection{Masking}

We mask out the Galactic plane using the field {\tt GAL040} from the non-apodized version of the HFI mask {\tt HFI\_Mask\_GalPlane-apo0\_2048\_R2.00} from {\em Planck}\footnote{\url{https://irsa.ipac.caltech.edu/data/Planck/release_2/ancillary-data/previews/HFI_Mask_GalPlane-apo0_2048_R2.00/index.html}}. This step masks most of the known X-ray sources in the Milky Way, the eROSITA bubbles that span the central part of the Milky Way, and the LMC and SMC. The eRASS1 map in the Eastern half of the Galactic hemisphere is not publicly available. 

Resolved point sources are masked out in the X-ray surface brightness map. We use the eRASS-DE DR1 main source catalog v1.1 \citep{eRASS1}. To select point sources, we apply the condition {\tt EXT\_LIKE}= 0, resulting in $903,521$ sources. For each point source, we used a masking radius of $2\times$ {\tt APE\_RADIUS\_1}, which is the radius of the aperture where the photons, in the 0.2-2.3~keV band from the source, are extracted from. 

We also mask out nearby galaxy clusters that contribute significantly to non-Gaussian covariances in the angular power spectrum \citep{osato_masahiro21}. We mask out regions within $R_{500c}$ of the clusters with $z \leq z_{\rm min} = 0.1$ selected from the eRASS1 galaxy cluster catalog {\tt erass1cl\_main\_v3.2} \citep{eRASS1_cluster_catalog}. The choice of $z_{\rm min} = 0.1$ is to maximize the signal-to-noise ratio of the angular power spectrum in the multipole range $\ell \in [100, 2000]$. \footnote{The signal-to-noise of the power spectrum increases by a factor of $\sim 5$ when we apply the redshift exclusion.}

This results in a total sky fraction of $f_{\rm sky} = 0.209$ that is unmasked. Figure~\ref{fig:eRASS1_map} shows the masked count rate maps used in our analysis. 

\subsection{Galactic Absorption}

To account for Galactic HI absorption, we use the full-sky map of HI column density from the HI4PI Collaboration \citep{HI4PI}, and compute the effective absorption for each pixel in our map using the {\tt tbabs} model \citep{tbabs} from XSPEC, assuming a flat spectrum in the $0.5-2$~keV band. The column density in the unmasked region is well described by the log-normal distribution, The mean and $\pm1\sigma$ scatter of $\log_{10}(N_{\rm HI}/{\rm cm^{-2}})$ is $20.4\pm 0.24$. To ensure that the angular variations due to HI absorption are taken care of, we divide the count rate by the absorption value on a pixel-by-pixel basis. 

\subsection{Measuring the Angular Power Spectrum}\label{sec:methods}

The first step to obtain the angular power spectrum is to measure the ``pseudo'' angular power spectrum, given by
\begin{equation}
    \tilde{C}_\ell = \frac{1}{2\ell+1}\sum^{\ell}_{m=-\ell} a_{\ell m}^*a_{\ell m},
\end{equation}
where $\ell$ is the multipole with $\ell \sim \pi/\theta$ for angular scale $\theta$ in radians. The term $a_{\ell m}$ is the coefficient of the spherical harmonics corresponding to the mode $(\ell,m)$ and $a_{\ell m}^*$ is the complex conjugate. 

The relationship between the pseudo angular power spectrum and $\tilde{C}_{\ell}$ and the true angular power spectrum ${C}_\ell$ is given by  
\begin{equation}
    \tilde{C}(\ell) = \sum_{\ell'} M(\ell,\ell') C(\ell')B(\ell')^2 + {N_\ell},
\end{equation}
where $M(\ell,\ell')$ is the mode coupling matrix, ${N_\ell}$ is the power spectrum due to shot noise, and $B(\ell')= \exp(-\ell(\ell+1)\sigma^2/2)$ is the beam transfer function, with $\sigma = \theta_{\rm FWHM} /\sqrt{8\ln 2}$ and $\theta_{\rm FWHM} = 25.7"$, corresponding to the pixel size of the HEALpix map with $N_{\rm side}=8192$. Note that $\theta_{\rm FWHM}$ is also approximately the PSF of eROSITA/SRG. 

We use the Python version of the {\tt NaMaster} code \citep{namaster, namaster2, namaster3} to compute $\tilde{C}_\ell$. We bin the power spectrum linearly with $(\ell_{\rm min}, \ell_{\rm max}) = (2, 24575)$, with bandpower $\Delta \ell = 40$. 
We decouple power at different multipoles due to masking and limited sky coverage with the mode coupling matrix $M(\ell,\ell')$ computed by the code. 

We estimate the shot noise term $N_\ell$ using the so-called `A-B' method \citep{cappelluti_etal13}. We divide the eRASS count rate map into two: map {A} with odd-numbered photons and map {B} with even-numbered photons. The difference of the two maps A-B should then contain only the background noise. We then measure the power spectrum of the differences of the two maps as the shot-noise power spectrum. 

\section{Model}\label{sec:model}

\subsection{Halo Power Spectrum}

We used the halo model to model the X-ray angular power spectrum. The details of the model can be found in \citet{shirasaki_etal19} and \citet{lau_etal23}. At a given angular scale $\ell$, the clustering power is given by
\begin{eqnarray}\label{eq:cl_icm}
C_{\ell} &=& C^{\mathrm{1h}}_{\ell} + C^{\mathrm{2h}}_{\ell}, \\
\label{eq:cl_1h}
C^{\mathrm{1h}}_{\ell} &=& \int_{z_{\rm min}}^{z_{\rm max}} dz\, 
\frac{dV}{dz} \nonumber \\
& \times & \int_{M_\mathrm{min}}^{M_\mathrm{max}} dM\, \frac{dn}{dM}
|S_\ell (M, z)|^2 ,  \\
\label{eq:cl_2h}
C^{\mathrm{2h}}_{\ell} &=& \int_{z_{\rm min}}^{z_{\rm max}} dz\, \frac{dV}{dz}  
P_L (k_\ell , z) \nonumber \\
& \times & \left[
\int_{M_\mathrm{min}}^{M_\mathrm{max}} dM\, \frac{dn}{dM}
b(M, z) S_\ell(M, z)
\right]^2,
\end{eqnarray}
where $k_\ell = \ell/\{ (1+z)d_A(z) \}$, $z_{\rm max}$ is the maximum redshift of observed clusters and groups, $d_A(z)$ is the angular diameter distance, $ dV/dz = (1+z)^2 d_A^2c/H(z)$ is the cosmological volume per redshift. Note that the one-halo term $C^{\mathrm{1h}}_{\ell}$ dominates over the two-halo term $C^{\mathrm{2h}}_{\ell}$ for $\ell \geq 100$. 
For the halo mass function $dn/dM$, and the halo bias function $b(M,z)$, we use those calibrated in \citet{tinker_etal08} and \citet{tinker_etal10}. For the linear matter power spectrum $P_L(k)$, we use the analytic approximation from \citet{eisenstein_hu98}. 

The lower and upper limits of the mass and redshift integrals are set to $(M_\mathrm{min},M_\mathrm{max}) = (10^{13}, 10^{15.5})\,M_\odot$, and $(z_{\rm min},z_{\rm max}) = (0.1, 2.0)$. The choice of $z_{\rm min}$ is to match the redshift cut we apply to the eRASS1 data to minimize non-Gaussian covariance. 

The X-ray surface brightness profile of a halo with mass $M$ and redshift $z$ in $\ell$-space is given by
\begin{equation}
\label{eq:S_ell}
S_\ell = \frac{4\pi R_s}{\ell_s^2}
\int dx\, x^2 S_X(x;z) \frac{\sin (\ell x/\ell_s)}{\ell x/\ell_{s}},
\end{equation}
where $x = r/R_s$, $\ell_{s} = d_A/R_{s}$, and $R_s$ is the scale radius (where the logarithmic slope of the underlying DM density profile is $-2$).
The term $S_X$ is the X-ray Surface Brightness profile in real space, in instrument units of photon counts per second per steradian,
\begin{eqnarray}\label{eq:xsb}
&& S_X(r;z) = \frac{n_H(r) n_e(r)}{4\pi(1+z)^4} \nonumber \\
&&\times \int^{E_{\rm max}(1+z)}_{E_{\rm min}(1+z)} \Lambda(T(r),Z,E) {\rm RMF}(E) {\rm ARF}(E) dE, 
\end{eqnarray}
where $n_H(r)$, $n_e(r)$ and $T(r)$ are profiles of the hydrogen and electron number densities and gas temperature respectively. 

We use the {\tt APEC} plasma code version 3.0.9 \citep{foster_etal12} to compute the X-ray cooling function $\Lambda$ and the energy conversion factor ECF = RMF $\times$ ARF that converts the flux to photon counts. The integrand is integrated over $[E_{\rm min}, E_{\rm max}] = [0.5, 2.0]$~keV in the observer's frame. We assume a constant metallicity of $Z=0.3 \,\rm{Z_\odot}$. For the ECF, we use the RMF file {\tt onaxis\_tm0\_rmf\_2023-01-17.fits.gz} and ARF file {\tt onaxis\_tm0\_arf\_filter\_2023-01-17.fits.gz}, downloaded from the eROSITA-DE DR1 website. \footnote{\url{https://erosita.mpe.mpg.de/dr1/eSASS4DR1/eSASS4DR1_arfrmf/}} We used the on-axis TM0 ARF because the exposure maps generated are already corrected for the number of detectors used as well as vignetting of the telescopes.

\subsection{Halo Gas Model}

The gas profiles are modeled using the Baryon Pasting (BP) model. We refer the reader to \citet{shaw_etal10, flender_etal17, osato_nagai23} for more details. The BP model has been well-tested and calibrated to match {\em Chandra} X-ray observations of clusters and groups.  Briefly, the total pressure (thermal plus non-thermal) of the halo gas is assumed to be in hydrostatic equilibrium with the halo potential of the underlying Dark Matter density (which is described by the NFW \citealt{nfw96} profile, and we use the halo concentration-mass relation (`all' halos) from \citealt{uchuu} ).  The gas density and the total gas pressure are related by the polytropic relation where
$P_{\rm tot} \propto \rho_{\rm gas}^\Gamma$, where the polytropic index $\Gamma$ is set to 1.2 \citep{shaw_etal10}.  The proportionality constants of the polytropic relations are determined by imposing energy and momentum conservation equations of the halo gas. The energy of the gas is given by
\begin{equation}\label{eq:energy}
E_{g,f}  =  E_{g,i} + E_{f} + \Delta E_p, 
\end{equation}
where $E_{g,f}$ and $E_{g,i}$ are the final and initial total energies (kinetic plus thermal plus potential) of the ICM. $\Delta E_p$ is the work done by the ICM as it expands.  The energy feedback from supernovae and active galactic nuclei $E_{f}$ is modeled as $E_{f} = \epsilon_f M_\star c^2$, where $\epsilon_f$ is the feedback efficiency parameter, which is treated as a free parameter in our model; and $M_\star$ is the stellar mass, where we adopt the best-fit $M_\star-M_{500c}$ relation in \citet[hereafter F17]{flender_etal17}. 

The solutions to the energy and momentum conservation equations depend on the boundary condition we imposed on.  We use the outer accretion shock radius $R_{\rm sh}$ as the boundary of the gas. We express $R_{\rm sh}$ in terms of the splashback radius $R_{\rm sp}$ and fit $R_{\rm sh}/R_{\rm sp}$ as a free parameter. For $R_{\rm sp}$, we use the fitting function in Equation~(7) of \citet{more_etal15b} calibrated from cosmological simulations 
\begin{equation}
R_{\rm sp}/R_{200m} = 0.81 \left(1+0.97 \exp{(-\nu_{200m}/2.44})\right),
\end{equation}
where $\nu_{200m} \equiv 1.686/\sigma(M_{200m},z)$, and $\sigma(M_{200m},z)$ is the variance of density fluctuation at the halo mass scale $M_{200m}$ at redshift $z$.
This sets our boundary condition for solving the conservation equations. Cosmological simulations suggest that $R_{\rm sh}/R_{\rm sp} \sim 2$ \citep{aung_etal21}, and we choose a uniform prior $[1.0, 3.0]$ around this value. 

Following F17, we model the gas in the cores of clusters and groups with a separate polytropic index $\Gamma_0$, which is smaller than the value in the outskirts $\Gamma=1.2$. The smaller $\Gamma_0$ model the core gas inside $<0.2R_{500c}$ as being cooler and denser, as observed. Lower $\Gamma_0$ results in higher X-ray surface brightness in the inner regions of clusters and groups, leading to higher X-ray power at smaller angular scales. We use the power spectrum to constrain the $\Gamma_0$ parameter. 

The temperature profile is computed by dividing the thermal pressure by the gas density. The thermal pressure is computed by subtracting the non-thermal pressure from the total pressure. We adopt the non-thermal pressure fraction profile from \citet{nelson_etal14}:
\begin{equation}
\frac{P_{\rm nt}(r)}{P_{\rm tot}(r)} =  1 - A_{\rm nt}\left[1+\exp\left\{-\left(\frac{r}{B_{\rm nt}\, R_{ 200m}}\right)^{C_{\rm nt}}\right\}\right], \label{eq:fnt}
\end{equation}
where $R_{\rm 200m}$ is the spherical over-density radius with respect to 200 times the mean matter density of the universe. The parameters are calibrated to be $A_{\rm nt} = 0.452$, $B_{\rm nt} = 0.841$, $C_{\rm nt} = 1.628$. 
The $R_{\rm 200m}$ scaling ensures halo redshift and mass independence at the cluster scales $M_{500c} \geq 3\times 10^{14} h^{-1}M_\odot$ over which this relation is calibrated with the {\em Omega500} cosmological simulation \citep{nelson_etal14b}. 
In our analysis, we take $A_{\rm nt}$, which represents the amplitude of the non-thermal pressure as the free parameter, and fixed other parameters to their fiducial values. 

Table~\ref{tab:params} summarizes the Baryon Pasting gas model parameters and their physical meaning. 

\begin{table*}
\centering
\begin{tabular}{|c|c|c|c|}
 \hline
 \textbf{Parameter} & \textbf{Physical Meaning} & \textbf{Prior} & \textbf{Posterior} \\
 \hline
 $10^6\epsilon_f$ & efficiency of feedback from AGN and SN & $\mathcal{U}(0.1, 20.0)$ & $11.73^{+1.39}_{-1.41}$\\
 $A_{\rm nt}$ & amplitude of the non-thermal pressure fraction & $\mathcal{U}(0.05, 0.95)$ & $0.47^{+0.13}_{-0.10}$ \\
 $\Gamma_0$ & polytropic exponent in halo cores ($<0.2 R_{500c}$)& $\mathcal{U}(0.0, 1.67)$ & $ > 0.93 $\\
 $R_{\rm sh}/R_{\rm sp}$ & halo gas boundary in unit of the splashback radius & $\mathcal{U}(1.0, 3.0)$ & $2.02^{+0.07}_{-0.12}$\\
 \hline
 $\Omega_M$ & matter density of the Universe & $\mathcal{U}(0.1, 0.5)$ & $0.40^{+0.02}_{-0.03}$ \\
 $\sigma_8$ & matter density fluctuation of the Universe & $\mathcal{U}(0.6, 1.0)$ & $0.69^{+0.04}_{-0.02}$ \\
 \hline
\end{tabular}
\caption{Parameters of the Halo Gas Model of the X-ray Angular Power Spectrum }
\label{tab:params}
\end{table*}

\subsection{Statistical Inference}

We perform statistical inference to the astrophysical and cosmological parameters, represented by the parameter vector $\theta = (\epsilon_f, A_{\rm nt}, \Gamma_0, R_{\rm sh}/R_{\rm sp}, \Omega_M, \sigma_8)$, where $\epsilon_f$, $A_{\rm nt}$, $\Gamma_0$, $R_{\rm sh}/R_{\rm sp}$ represent the feedback efficiency, the amplitude of non-thermal pressure fraction, and the polytropic index in the halo core ($<0.2 R_{500c}$), the ratio of the halo gas boundary (i.e. the outer accretion shock radius) to the splashback radius, respectively, and $\Omega_M$ and $\sigma_8$ are the matter density of the Universe and the amplitude of the matter density power spectrum. We infer the best-fit parameters to the eRASS1 power spectrum assuming a Gaussian likelihood function:
\begin{equation}
\ln \mathcal{L} = - \frac{1}{2}(C_\ell^{\rm data }- C_\ell^{\rm mod}(\theta)) \mathcal{C}^{-1} (C_\ell^{\rm data }- C_\ell^{\rm mod}(\theta))^T,
\end{equation}
where $C_\ell^{\rm data }$ and $C_\ell^{\rm mod}(\theta)$ are the measured and model angular power spectra, respectively, and $\mathcal{C}$ is the covariance matrix. We use the {\tt emcee} package \citep{emcee} to perform a Monte Carlo Markov Chain (MCMC) to sample the posterior distributions of the parameters with a total of 5000 steps with 48 walkers. We assume uniform flat priors on all parameters: $10^6\epsilon_f \in [0.1,10.0]$, $A_{\rm nt} \in [0.05, 0.85]$, $\Gamma_0 \in [0.0, 1.67]$, $R_{\rm sh}/R_{\rm sp} \in [1.0, 3.0]$, $\Omega_M \in [0.2, 4.0]$, $\sigma_8 \in [0.6, 1.0]$. 
Table~\ref{tab:params} summarizes the parameters we are fitting for and their priors. 

For the covariance matrix $\mathcal{C}(\ell, \ell')$, we separate it into Gaussian $\mathcal{C}^{\rm G}$ and non-Gaussian $\mathcal{C}^{\rm NG}$ components, with $\mathcal{C} = \mathcal{C}^{\rm G}+\mathcal{C}^{\rm NG}$ . For the Gaussian part, we estimate using the analytical method {\tt gaussian\_covariance} provided in {\tt NaMaster}. 

For the non-Gaussian part, we estimate it as 
\begin{equation}   
\mathcal{C}^{\rm NG}(\ell, \ell') = \frac{1}{4\pi f_{\rm sky}} T(\ell, \ell'),
\end{equation}
where $T(\ell, \ell')$ is the halo model trispectrum,
\begin{eqnarray}
T(\ell, \ell') &=& \int_{z_{\rm min}}^{z_{\rm max}} dz\,
\frac{dV}{dz} \nonumber \\
& \times & \int_{M_\mathrm{min}}^{M_\mathrm{max}} dM\, \frac{dn}{dM}
|S_\ell (M, z)|^2 |S_{\ell'} (M, z)|^2.
\end{eqnarray}
with the same redshift and halo mass range as the power spectrum. Note that we ignore the two-halo term in the trispectrum, since it is negligible for the multipole $\ell \geq 100$ we are probing. 
We have also ignored the Super-Sample Covariance \citep[SSC,][]{takada_hu13}. Using the halo model approximations of SSC \citep{osato_masahiro21}, we have verified that the SSC terms are much smaller than the covariance from both the Gaussian part and the trispectrum, by $\sim 100$ for the multipole range we probe. Hence, we do not include SSC in the covariance matrix.

\begin{figure}[htbp]
    \centering
    \includegraphics[width=0.475\textwidth]{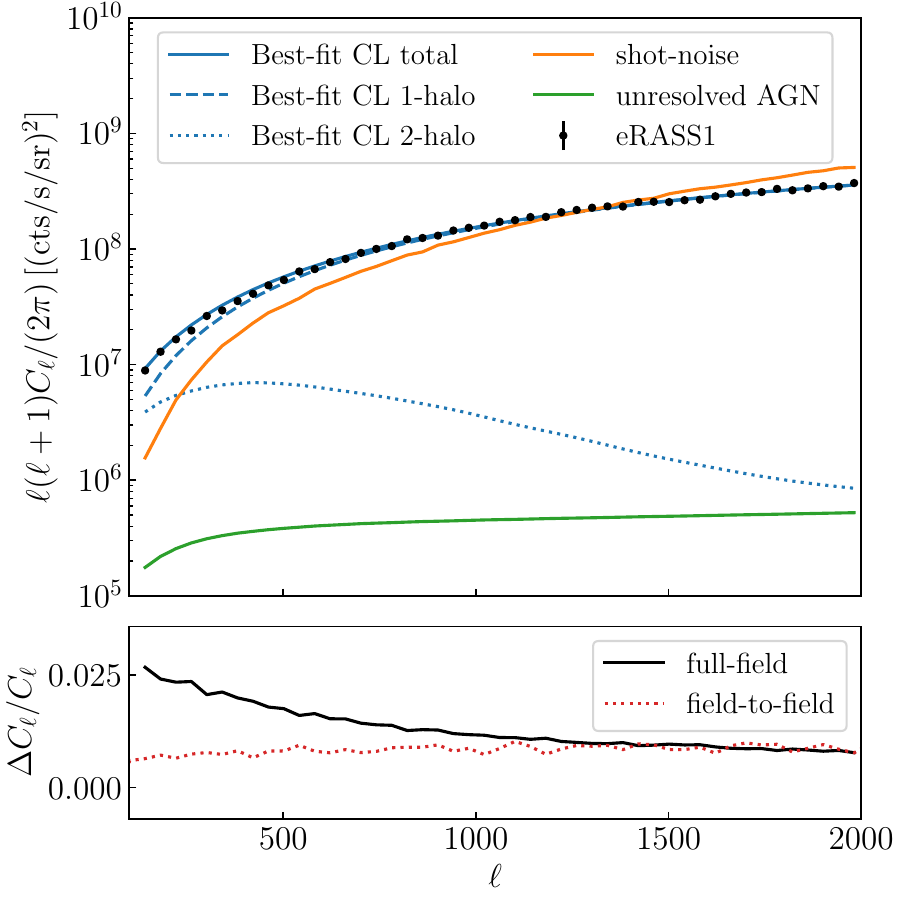}
    \caption{{\em Top} panel: The black data points show the X-ray angular power spectrum signal measured on the masked eRASS map with shot-noise power spectrum (orange line) subtracted. The blue solid line shows the best-fit model of the power due to clusters and groups. We also show the break-down of best-fit model into the one-halo (blue dashed line) and two-halo (blue dotted line) terms.   The estimated angular power spectrum due to unresolved AGN is shown in green. {\em Bottom} panel shows the fractional errors of the power spectrum measurement, computed as the $1\sigma$ error divided by the power spectrum signal (black line). We compare this to the `field-to-field' jackknife fractional error estimate (red dashed line), which is computed from 12 power spectra each measured by dropping one of the 12 equally-divided regions from the  footprint, one at a time. This `field-to-field' error is smaller (comparable) at low (high) multipoles than the fiducuial `full-field' error that we used in our statistical inferences. } 
    \label{fig:eRASS1_power}
\end{figure}

To test the covariance calculation, we adopt a jackknife approach by dividing our footprint into 12 regions, and computing the power spectrum by removing one region at a time. We then compute the covariance matrix with the 12 power spectra. We refer this covariance estimate as `field-to-field' covariance. We found that this `field-to-field' estimate to be smaller at low $\ell$ and comparable to the fiducial `full-field' covariance at high $\ell$. Thus, by using the fiducial `full-field' covariance in our statistical inference, we are being conservative in the statistical constraints of the parameters. The comparison of their diagonal components is shown in the bottom panel in Figure~\ref{fig:eRASS1_power}. 

\section{Results}
\label{sec:results}

\begin{figure*}[htbp]
    \centering
    \includegraphics[width=2.0\columnwidth]{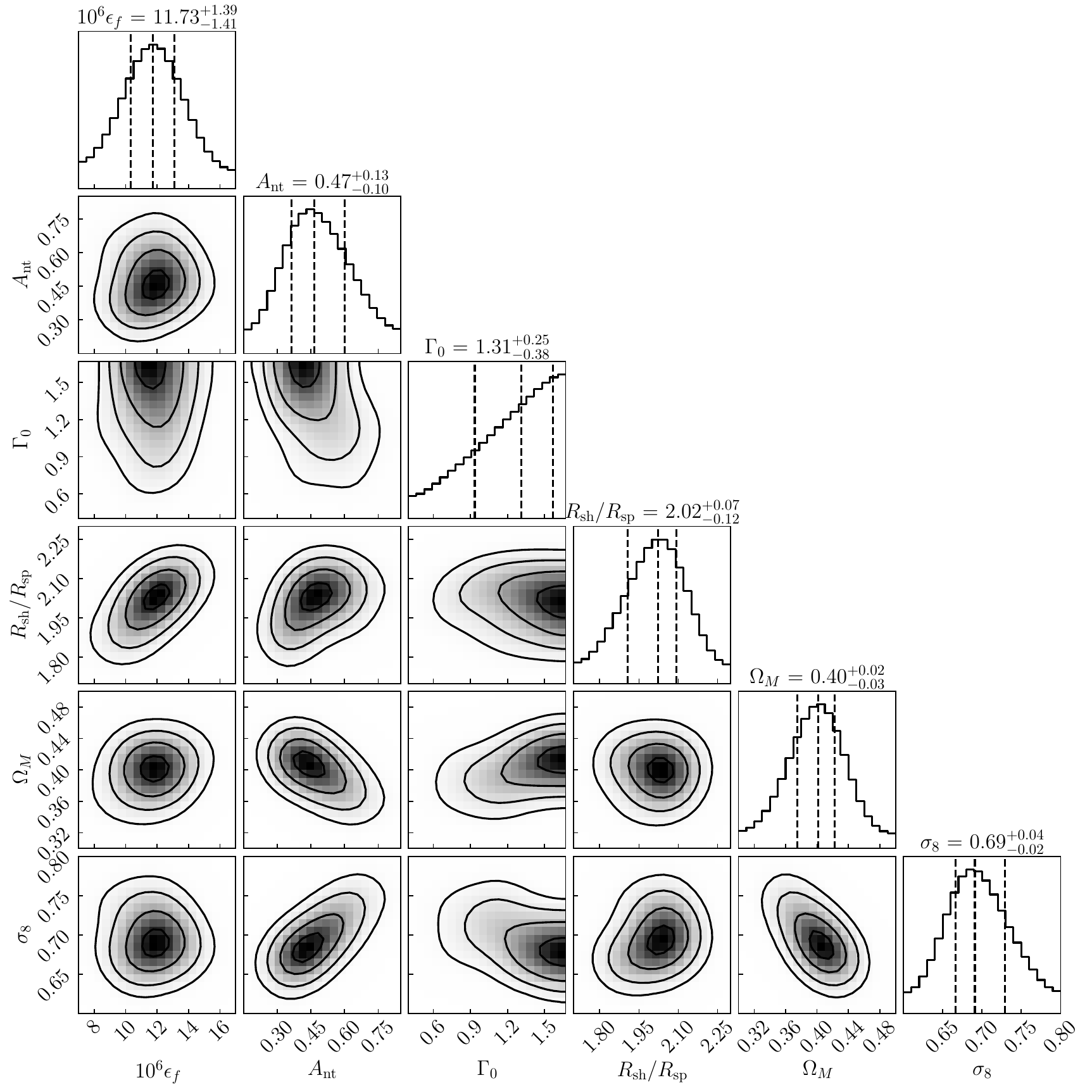}
    \caption{Posterior distributions of the parameters of our halo model when comparing to the eRASS1 power spectrum. The contours show $1,2$ and $3\sigma$ uncertainties for the parameters. The quoted errors on the top of each column represent the 68\% confidence interval.  }
    \label{fig:eRASS1_mcmc}
\end{figure*}

\begin{figure}[htbp]
    \centering
    \includegraphics[width=0.45\textwidth]{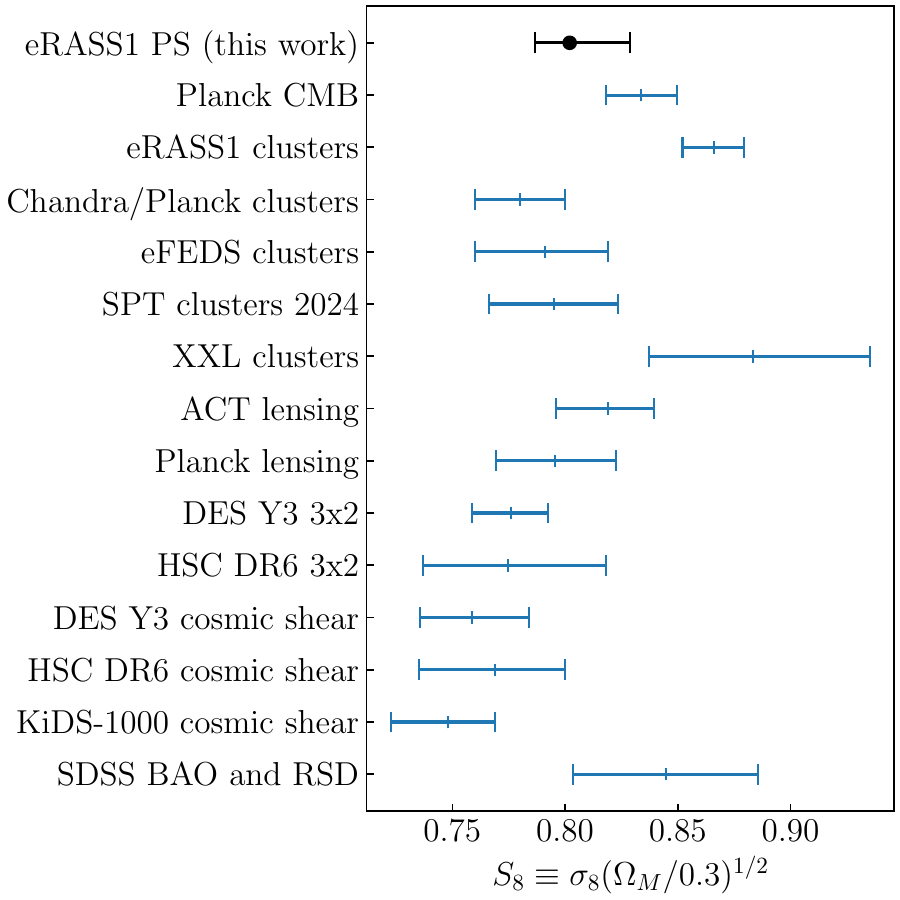}
    \caption{Comparison of the $S_8$ constraint from the eRASS1 angular power spectrum with the constraints obtained from other cosmological probes in the literature: {\em Planck} CMB \citep{planck2018}, cluster abundances from eRASS1 \citep{eRASS1_cluster_abund}, {\em Planck} clusters with {\em Chandra}-derived cluster masses \citep{chandra_planck24}, SPT \citep{spt24}, eFEDS \citep{efeds23}, and XXL \citep{xxl22}; CMB lensing from ACT \citep{act_lensing24a, act_lensing24b}, and Planck \citep{planck_lensing20}; Galaxy clustering and weak lensing ($3\times2$) from DES Y3 \citep{des_y3_3x2pt22}, HSC DR6 \citep{hsc_3x2pt23a, hsc_3x2pt23b}, and KiDS-1000 \citep{kids_3x2pt21}; Cosmic Shear from DES Y3 \citep{des_y3_cosmicshear22}, HSC DR6 \citep{hsc_cosmicshearps23, hsc_cosmicshear_2pcf23}, and KiDS-1000 \citep{kids_cosmicshear23}; and SDSS Baryon Acoustic Oscillation (BAO) plus Redshift Space Distortion (RSD) \citep{sdss_bao_rsd21}. 
    The error bars represent $68\%$ confidence level.  }
    \label{fig:s8_comparison}
\end{figure}

\subsection{Power Spectrum}

Figure~\ref{fig:eRASS1_power} shows the X-ray angular power spectrum measured with resolved point sources and the masked Galactic plane, with the shot-noise power subtracted. We also show the shot-noise power spectrum and the expected power due to unresolved AGN. The power spectrum of the unresolved AGN is estimated using the model described in Appendix~\ref{sec:agn_hod}, calibrated using the eFEDS measurements from \cite{comparat_etal23}. Both the shot noise and the unresolved AGN power are subdominant to the power of clusters and groups. For inference, we only use the multipole range $\ell = [100, 2000]$ which is robust to large-scale fluctuations and the shot-noise power. 

We also show the one-halo and two-halo terms of the best-fit model power spectrum in Figure~\ref{fig:eRASS1_power}. It shows that the one-halo term is significantly larger than the two-halo term in the multipole range $\ell = [100, 2000]$ explored. The dominance of the one-halo term in this multipole range is consistent  with the predictions for the thermal SZ power spectrum \citep[e.g.][]{komatsu_seljak02, hill_pajer13, bolliet_etal18}. Thus, the one-halo term provides most of the  cosmological and astrophysical constraining power of the X-ray angular power spectrum. 

\subsection{Cosmological Constraints}

Figure~\ref{fig:eRASS1_mcmc} shows the posterior distributions resulting from MCMC. For cosmological parameters, the marginalized constraints are $\Omega_M=0.40^{+0.02}_{-0.03}$ and $\sigma_8=0.69^{+0.04}_{-0.02}$, resulting in $S_8=0.80^{+0.02}_{-0.01}$, (the quoted errors are all $1\sigma$ uncertainties unless stated otherwise). Figure~\ref{fig:s8_comparison} shows the comparison against values reported from different cosmological probes, including {\em Planck} CMB \citep{planck2018}, and other LSS measurements. Our value is lower than {\em Planck} CMB constraints \citep{planck2018} with $S_8 = 0.83 \pm 0.02$, but they are marginally consistent with $\sigma$.  Our $S_8$ constraint is also consistent with other Large-Scale Structure probes with a similar level of uncertainties, except eRASS1 cluster abundance and XXL cluster abundance, which report higher $S_8$ that are in tension with our results at $>1\sigma$. 

Our best-fit $\Omega_M=0.40^{+0.02}_{-0.03}$ value is higher than the {\em Planck} CMB value, and our best-fit $\sigma_8=0.69^{+0.04}_{-0.02}$ value is lower than the {\em Planck} CMB value. However, we also notice that our $\Omega$ and $\sigma_8$ constraints are highly degenerate with each other, as well as with other astrophysical parameters in our model. 

\subsection{Astrophysical Constraints}

The eRASS1 power spectrum provides constraints on the astrophysics parameters: the feedback efficiency $\epsilon_f$, the non-thermal pressure fraction amplitude $A_{\rm nt}$, the polytropic index in cluster core $\Gamma_0$, and the gas halo boundary $R_{\rm sh}/R_{\rm sp}$. 
The best-fit value for the feedback is $10^6\epsilon_f = 11.73^{+1.39}_{-1.41}$, which is significantly higher than the best-fit value of $10^6\epsilon_f = 3.97^{+4.82}_{-2.88}$ from F17, which uses the halo model to fit the density profiles from the {\em Chandra} galaxy cluster sample selected via the SZ effect with the South Pole Telescope. However, there is an important difference between the model used in F17. In F17, the halo boundary is fixed at the virial radius, while in the present work we fit for the halo boundary. Changing the halo boundary in the gas model induces changes in both the shape and normalization of the X-ray emissivity profile, thus impacting the feedback parameter derived from it. 

For the non-thermal pressure fraction amplitude, we obtain an upper limit on $A_{\rm nt} = 0.47^{+0.13}_{-0.10}$. Our constraints imply a consistent non-thermal pressure fraction compared to those predicted from non-radiative cosmological hydrodynamical simulations, which has $A_{\rm nt} = 0.45$ \citep{nelson_etal14b}. This is different from previous statistical constraints obtained from tSZ-Weak lensing cross-correlations between {\em Planck} and the Red Cluster Sequence Lensing Survey \citep{osato_etal18}, which implied around 2 times larger non-thermal pressure fraction compared to predictions from cosmological simulations.  

The power spectrum eRASS1 constrains the polytropic index $\Gamma_0$ of gas in the cores ($R<0.2R_{500c}$) of the clusters and groups with a lower limit $\Gamma_0 > 0.93$. This value is higher compared to the best fit of $0.10$ from {\em Chandra}-SPT cluster profiles in F17. This lower limit value is consistent with the observed polytropic index $\Gamma \approx 1.2$ outside the cluster core \citep{ghirardini_etal19}. 

The eRASS1 power spectrum also constrains the outer boundary of the halo gas to be larger than the splashback radius by $R_{\rm sh}/R_{\rm sp} = 2.02^{+0.07}_{-0.12}$ times the splashback radius. This is consistent with the prediction from the cosmological simulation $R_{\rm sh}/R_{\rm sp}=1.89$ \citet{aung_etal21}, and measurements with the stacked thermal SZ profile of clusters observed with the South Pole Telescope $R_{\rm sh}/R_{\rm sp} > 2.16\pm 0.59$ \citep{anbajagane_etal22}. 

The best-fit cosmological and astrophysical parameters are summarized in Table~\ref{tab:params}. 

\section{Discussion}
\label{sec:discussion}
In this section, we discuss probable systematic uncertainties and their potential impact on the cosmological and astrophysical constraints that we obtained from the eRASS1 X-ray power spectrum. 

\subsection{Contamination and Mis-classification}

The current analysis assumes the complete removal of resolved point sources. Any residual point sources can contaminate our measurement. We rely on the eRASS1 main source catalog \citep{eRASS1} to provide a list of resolved point sources. However, our power spectrum can be contaminated by resolved point sources that we fail to mask out. This will happen if the point source is misclassified as an extended source. According to mock eRASS1 simulations \citet{seppi_etal22}, it is estimated that 25\% of AGN are likely to be classified as extended sources. These potentially misclassified clusters and groups are likely to be very small in angular size, either due to their low masses or at high redshifts. Otherwise, they are unlikely to classified as point sources if they were more extended. These high-redshift and low mass clusters and groups contribute a small fraction $<1\%$ to the overall diffuse power spectrum \cite[see Figure~1 in][]{shirasaki_etal19}.  Furthermore, a change in clustering power due to mis-classified AGN by e.g., $\sim 20\%$ only leads to relatively small change in $\sigma_8$ of $\sim 5\%$ \citep[e.g. Figure~3 in][]{lau_etal23}. 
On the other hand, because of the relatively large PSF of eROSITA, compact groups and clusters can be mis-identified as point sources \citep[e.g.][]{bulbul_etal22}, but only around 4\% \citep{seppi_etal22}.  By limiting the multipole range to $\ell \leq 2000$, we exclude potential point source contaminants. 

Another potential source of systematic error is the fluctuations in Galactic HI absorption. 
To estimate its impact, we computed the angular power spectrum of the Galactic HI absorption. For the multipole range we used in our analysis, the clustering amplitude due to $N_{\rm HI}$ absorption is very small, from $\sim 10^{-4}$ at $\ell =100$ to $\sim 10^{-6}$ at $\ell=2000$. Thus we believe that this systematic uncertainty has negligible impact on our results. 

\subsection{Impact of Gas Clumping}

There are also a few limitations in our power spectrum model. First, gaseous substructures in halos can potentially boost the X-ray emissions via gas density clumping \citep{mathiesen_etal99, nagai_lau11}. Specifically, the amplitude of the X-ray power spectrum can be boosted by the square of the gas density clumping factor $C \equiv \langle n_{\rm gas}^2\rangle/\langle n_{\rm gas}\rangle^2 \geq 1$. Observations of nearby galaxy clusters indicate that the gas density clumping is non-zero, with clumping factor $C < 1.21$ from observations of nearby clusters \citep{eckert_etal15, mirakhor_walker21}. However, current estimations of gas clumping have large uncertainties. The measurements are also limited to nearby massive clusters where selection biases are uncertain. Thus, we refrain from assuming clumping in our analysis.  The clumping factor can be constrained in the future with cross-angular power spectrum in other wavelengths, such as X-ray/SZ/Lensing cross-power spectra \citep{shirasaki_etal19}.  Second, we ignore the scatter in the gas profiles of the halos in the halo model. Such scatter adds additional fluctuations in the X-ray surface brightness, and thus ignoring the scatter can lead to underestimates of the clustering signal. The impact of this effect on our results is uncertain and we will study it in future works. 

\begin{figure}[htbp]
    \centering
    \includegraphics[width=0.5\textwidth]{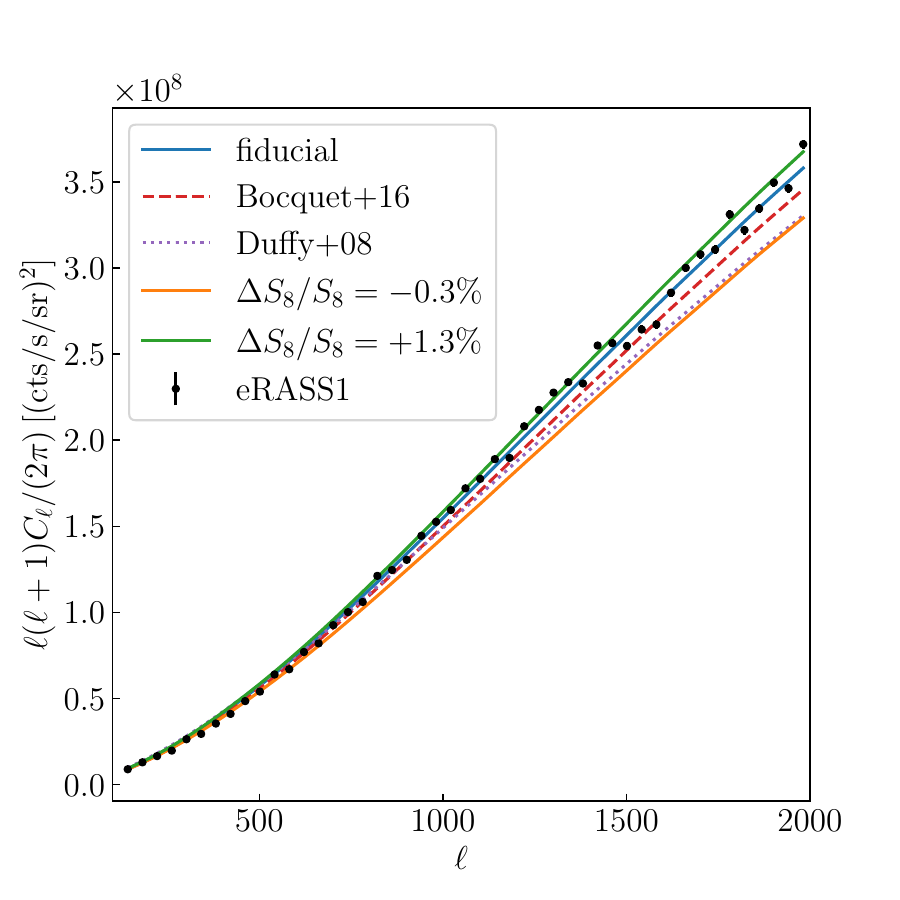}
    \caption{ We compare the model power spectra computed with a different halo mass function from \citet{bocquet_etal16} (red dashed line), with a different halo concentration mass relation from \citet{duffy_etal08} (purple dotted line), from our fiducial model with the halo mass function from \citet{tinker_etal08} and the halo concentration relation from \citet{uchuu} (blue line). Note that the model power spectra above are computed with the best-fit values. We also show the fiducial model power spectra computed with $S_8$ that are $1.3\%$ larger (green solid line) and $0.3\%$ smaller (orange solid line) than the best-fit value, well-within the uncertainties of $S_8$ obtained from MCMC. The black dots are the eRASS1 power spectrum data points, same as those shown in Figure~\ref{fig:eRASS1_power}. 
    }
    \label{fig:hmf_cm}
\end{figure}

\subsection{Systematics due to Halo Mass Function and Halo Concentration-Mass Relation}

In our analysis, we chose specific models for the halo mass function and the halo concentration relation, namely, those from \citet{tinker_etal08} and \citet{uchuu}, respectively. Such modeling choices could be biased and lead to additional systematic uncertainties in our cosmological and astrophysical constraints. We investigate potential systematic uncertainties due to the assumed halo mass function and the halo concentration models. Figure~\ref{fig:hmf_cm} shows the model power spectrum computed with a different halo mass function from \citet{bocquet_etal16}, and the power spectrum computed with the halo concentration-mass relation from \citet{duffy_etal08}. These power spectra are computed using the same best-fit parameters in Table~\ref{tab:params}. In the same figure, we also show the power spectra with $S_8$ that is $1.3\%$ larger and $0.3\%$ smaller than the best-fit value. It shows that changing the underlying models of the halo mass function and the halo concentration-mass relation leads to smaller changes in the power spectrum compared to changing $S_8$ by $+1.3\%$ and $-0.3\%$. Thus, the systematic uncertainties due to modeling choices in halo mass function and halo concentration,  is limited to $\Delta S_8/S_8 = +1.3\%, -0.3\%$, which is smaller than the posterior uncertainties $\Delta S_8/S_8 = +2.5\%, -1.3\%$ obtained from MCMC.

\subsection{Comparison with eRASS1 Measurements of Clustering of Clusters in Seppi et al.~(2024)}

Finally, we compare our results with recent clustering measurements of  of eRASS1 galaxy clusters with the  two-point correlation function (2PCF) from \citet{eRASS1_cl_clustering}. They constrain $\sigma_8 = 0.76^{+0.22}_{-0.16}$, with relatively large uncertainties compared to ours. This can be explained due to two key differences. First, the 2PCF measurement was performed on galaxy clusters that are detected in the eRASS1 \citep{eRASS1_cluster_catalog}, counting cluster pairs in the catalog. For the angular power spectrum, we directly measured the clustering of X-ray photons coming from all clusters and groups, including those that are undetected or unresolved in the eRASS1 cluster catalog. Since statistical uncertainties of any two-point statistics scales inversely with the square-root of the number of pairs,  the X-ray angular power spectrum has smaller statistical uncertainties than the 2PCF measurements as there are more photon pairs than cluster pairs. Furthremore, unlike the 2PCF in \citet{eRASS1_cl_clustering} which the two-halo term dominates (as 7 out of their 9 clustering data correspond to scales $>2 h^{-1}{\rm Mpc}$) and whose normalization depends linearly on $\sigma_8^2$, the X-ray angular power spectrum measures mostly the one-halo term (see Figure~\ref{fig:eRASS1_power}) which dependence on the halo mass function (Equation~\ref{eq:cl_1h}) makes it very sensitive (nearly exponential) to $S_8$. These key differences explain the different $S_8$ constraints between our angular power spectrum measurements and the 2PCF measurements from \citet{eRASS1_cl_clustering}. 

\section{Conclusions}
\label{sec:summary}

In this paper, we present the first constraints on the cosmological parameter $S_8 \equiv \sigma_8(\Omega_M/0.3)^{1/2}$ based on the X-ray angular power spectrum of clusters and groups measured from the half-sky map of eROSITA All Sky Survey Data Release 1 (eRASS1) in the energy band $0.5-2.0$~ keV.  We obtain marginalized constraints on $S_8 = 0.80^{+0.02}_{-0.01}$. Our constraint is consistent with those inferred from the Large-Scale Structure probes at low redshift. Our constraint is lower than that of the {\em Planck} CMB primary anisotropy measurements \citep{planck2018} to within $1\sigma$. 

From the eRASS1 power spectrum, we also obtain constraints on the cluster astrophysics parameters. Specifically, we find that the eRASS1 power spectrum prefers stronger feedback and higher polytropic index,  compared to those inferred from X-ray observations of SZ-selected cluster profiles in F17, and a similar level of non-thermal pressure compared to predictions of cosmological simulations.   

Our work demonstrates that the X-ray angular power spectrum of galaxy clusters and groups is a viable and competitive probe of cosmology. There are a few areas that one can pursue to improve cosmological constraints. First, the major systematic uncertainties in the constraints of the current work are the degeneracies between astrophysical and cosmological parameters. These degeneracies can be broken down with better constraints on the feedback physics, non-thermal pressure, cool-core properties, and gas clumping in galaxy clusters and groups. Measurements of the thermodynamic properties of clusters and groups (such as profile measurements and scaling relations between global thermodynamical variables and halo mass) with {\em Chandra} and XMM-{\em Newton}, as well as with thermal SZ and cosmic shear measurements \citep[e.g.,][]{laposta_etal24} will be useful for constraining these astrophysics of clusters, thus improving the cosmological constraints by breaking the degeneracies. Their higher angular resolution will also help mitigate contamination from clustering of point sources. 

Second, eRASS map with longer exposure will help reduce statistical uncertainties in the current power spectrum measurements. With the future release of the 4 complete sky scans from eRASS4, with $\sim 4$ times deeper exposure, the statistical error on the power spectrum will be reduced by a factor of $\sim 2$, due to suppression of the shot-noise power spectrum. 

Finally, by cross-correlating the X-ray angular power spectrum with thermal SZ and lensing power spectra with future large-scale multiwavelength surveys such as CMB-S4, Rubin and Euclid Observatories in the optical, we expect to yield tight constraints on both cluster astrophysics and cosmology \citep{shirasaki_etal19}. In addition to $S_8$, we will also be able to constrain other cosmological parameters, such as the Dark Energy equation of state to percent-level precision with X-ray, SZ, and weak lensing cross-power spectra.

\section*{Acknowledgments}
We thank the referee for useful comments and feedback. 
We thank Jeremy Sanders and Miriam Ramos-Ceja for their advice on the eRASS1 half-sky map construction. 
E.T.L. acknowledges support from the Chandra grant AR4-25014X. 
\'A.B. acknowledges support from the Smithsonian Institution and the Chandra Project through NASA contract NAS8-03060. D.N. is supported by NASA ATP23-0154 grant. 
M.S. acknowledges support from MEXT KAKENHI Grant Number (20H05861, 23K19070, 24H00215, 24H00221).
This work is supported by the Yale Center for Research Computing facilities and staff. 
This work is based on data from eROSITA, the soft X-ray instrument on board SRG, a joint Russian-German science mission supported by the Russian Space Agency (Roskosmos), in the interests of the Russian Academy of Sciences represented by its Space Research Institute (IKI), and the Deutsches Zentrum f\"ur Luft- und Raumfahrt (DLR). The SRG spacecraft was built by Lavochkin Association (NPOL) and its subcontractors and is operated by NPOL with support from the Max Planck Institute for Extraterrestrial Physics (MPE). The development and construction of the eROSITA X-ray instrument was led by MPE, with contributions from the Dr.\ Karl Remeis Observatory Bamberg \& ECAP (FAU Erlangen-Nuernberg), the University of Hamburg Observatory, the Leibniz Institute for Astrophysics Potsdam (AIP), and the Institute for Astronomy and Astrophysics of the University of T\"ubingen, with the support of DLR and the Max Planck Society. The Argelander Institute for Astronomy of the University of Bonn and the Ludwig Maximilians Universit\"at Munich also participated in the science preparation for eROSITA. The eROSITA data shown here were processed using the eSASS software system developed by the German eROSITA consortium.

\facilities{eROSITA/SRG}

\software{
eSASS \citep{eFEDS},
healpy \citep{healpy},
pymaster \citep{namaster},
emcee \citep{emcee}
}

\appendix

\section{Modeling the Angular Power Spectrum due to Unresolved AGN}
\label{sec:agn_hod}

Following \citet{helgason_etal14, singh_etal17, comparat_etal23}, we model the contribution of power due to the clustering of unresolved AGNs using the Halo Occupation Distribution (HOD) model and the AGN X-ray luminosity function (XLF). The mean number of AGN residing in given dark matter halo with given mass $M$ is described by the HOD model from \citet{more_etal15} as :
\begin{eqnarray}\label{eq:agn_hod}
 N(M) &= &N_c + N_s, \\
 N_c(M) &=& \frac{1}{2}\left [ 1+{\rm erf}\left (\frac{\log_{10} M - \log_{10} M_{\rm min}}{\sigma_{\log_{10} M}} \right ) \right ], \\
  N_s(M) &=& N_c \left(\frac{M -10^{M_{\rm sat}-1}}{10^{M_{\rm sat}}} \right)^{\alpha},
\end{eqnarray}
where $N_c$ and $N_s$ are the mean numbers of central AGN satellite AGN, respectively. We adopt the HOD parameters from \citet{comparat_etal23} which are derived from the model fitted with the eFEDS AGN clustering measurements. 
The XLF $\Phi_{\rm AGN} \equiv dn_{\rm AGN}/{d\ln L_X}$ describes the average comoving number density of AGN $n_{\rm AGN}$ at a given redshift $z$ per natural log of AGN X-ray luminosity $\ln L_x$. We use the LADE XLF model in the 0.5-2~keV band \citep{aird_etal15}:
\begin{equation}\label{eq:xlf}
    \Phi_{\rm AGN}(L_X, z) = K(z) \left[ \left(\frac{L_X}{L_*(z)}\right)^{\gamma_1} + 
  \left(\frac{L_X}{L_*(z)}\right)^{\gamma_2} \right]^{-1}, 
\end{equation}
where 
\begin{eqnarray}
    K(z) &=&K_0 \times 10^{d(1+z)}, \\
    L_*(z)&=&L_0 \left[ \left(\frac{1+z_c}{1+z}\right)^{p_1}+\left(\frac{1+z_c}{1+z}\right)^{p_2} \right ]^{-1},    
\end{eqnarray}
The HOD and XLF model parameters are summarized in Table~\ref{tab:agn_hod}.

\begin{table}
\centering
\begin{tabular}{|l|c|}
 \hline
 \textbf{Parameter} & \textbf{Value}  \\
 \hline
  $\log_{10}(M_\mathrm{min})$ & 13.09 \\
  $\sigma_{\log_{10} M}$ & 1.3 \\
  $M_{\rm sat}$ & 14.65 \\
  $\alpha$ & 0.75 \\
 \hline
 $\log K_0 [\mathrm{Mpc^{-3}}]$ & -4.28 \\
 $\log L_0 [\mathrm{erg\,s^{-1}}]$ & 44.93  \\
 $\gamma_1$ & 0.44 \\
 $\gamma_2$ & 2.18 \\
 $p_1$ & 3.39\\
 $p_2$ & -3.58  \\
 $z_c$ & 2.31 \\
 \hline
\end{tabular}
\caption{Parameters of the AGN clustering model. }
\label{tab:agn_hod}
\end{table}

The one- and two-halo terms of the AGN clustering is then given by
\begin{eqnarray}
C^{\rm 1h}_{\rm AGN}(\ell) &=& \int {\rm d}z \frac{dV}{dz}  W_{\rm AGN}(z)
    \int dM \frac{dn}{dM} \nonumber \\
&\times& \left( \frac{N_s^2 u_{\rm AGN}^2 + 2N_c N_s u_{\rm AGN}}{n_{\rm AGN}^2}\right), \\
    C^{\rm 2h}_{\rm AGN}(\ell) &=& \int dz \frac{dV}{dz} P_{L}(k,z)  W_{\rm AGN}(z)\nonumber \\
&\times& \left[ \int dM \frac{dn}{dM} b(M,z)\left(\frac{ N_c+ N_s u_{\rm AGN}}{n_{\rm AGN}}\right ) \right]^2,   
\end{eqnarray}
where 
\begin{equation}
W_{\rm AGN}(z) = \int d\ln L_X \Phi_{\rm AGN}(L_X, z) [S_X(L_X, z)]^2 f(S_X),
\end{equation}
with $S_X$ being the AGN flux in ${\rm erg\,s^{-1}\,cm^{-2}}$, and
\begin{equation}
u_{\rm AGN}(k=\ell/\chi;M,z) = \int dr 4\pi r^2 \tilde{u}_{\rm AGN}(r; M, z) \frac{\sin(kr)}{kr},
\end{equation}
is the Fourier transform of the normalized radial distribution $\tilde{u}_{\rm AGN}$ of AGN inside halo of mass $M$, at redshift $z$, and at comoving distance $\chi$. We assume $\tilde{u}_{\rm AGN}$ follows the NFW profile with concentration following the relation from \citet{uchuu}, and is normalized such that its volume integral within the virial radius is unity. We model the flux selection curve $f(S_X)$ as a logistic function with
\begin{equation}
f(S_X) = \frac{1}{1+\exp(x)}, \;{\rm where} \; x = \frac{\log_{10}S_X-\log_{10}S_{\rm lim}}{\log_{10} \sigma_{\rm AGN}},
\end{equation}
and $\log_{10} (S_{\rm lim}/{\rm erg\,s^{-1}\,cm^{-2}}) = 13.5$ and $\log_{10}\sigma_{\rm AGN} = 0.12$ are set
to match the published eRASS1 flux limit \citep{eRASS1}. 

The kernel $W_{\rm AGN}(z)$ is integrated over the $L_X(z)$ for the range that corresponds to the k-corrected flux range $S_X/{\rm erg\,s^{-1}\,cm^{-2}} \in [10^{-20}, 10^{-10}] $ at the given redshift. We assume an energy spectral index of $\Gamma_{\rm AGN}=1.7$. We checked that the results do not change significantly for $\Gamma_{\rm AGN}=1.5, 1.9$. 

We note that the clustering power due to unresolved AGN contribution is always subdominant to the contributions from clusters, groups, and diffuse IGM. Thus, our results are insensitive to the parameters adopted of the AGN model in Table~\ref{tab:agn_hod}.

\bibliography{references}
\bibliographystyle{aasjournal}

\end{document}